\documentclass[12pt,aps,floats,showpacs]{revtex4}
\usepackage{graphicx}
\usepackage{epsfig}

\newcommand{\be}{\begin{equation}}
\newcommand{\ee}{\end{equation}}
\newcommand{\bea}{\begin{eqnarray}}
\newcommand{\eea}{\end{eqnarray}}

\begin{document}
\title{Classical dynamics of quantum fluctuations}

\author{Ram Brustein and David H. Oaknin}

\address{
 Department of Physics, Ben-Gurion University,
Beer-Sheva 84105, Israel\\
email: ramyb,doaknin@bgumail.bgu.ac.il}

\begin{abstract}

It is shown that the vacuum state of weakly interacting quantum
field theories can be described, in the Heisenberg picture, as a
linear combination of randomly distributed incoherent paths that
obey classical equations of motion with constrained initial
conditions. We call such paths ``pseudoclassical" paths and use
them to define the dynamics of quantum fluctuations. Every
physical observable is assigned a time-dependent value on each
path in a way that respects the uncertainty principle, but in
consequence, some of the standard algebraic relations between
quantum observables are not necessarily fulfilled by their
time-dependent values on paths. We define ``collective
observables" which depend on a large number of independent
degrees of freedom, and show that the dynamics of their quantum
fluctuations can be described in terms of unconstrained classical
stochastic processes without reference to any additional external
system or to an environment. Our analysis can be generalized to
states other than the vacuum.
Finally, we compare our formalism to the formalism of coherent
states, and highlight their differences.

\end{abstract}

\maketitle

\section{Introduction}

It is well known that some effects of quantum fluctuations of
commuting observables in a stationary state of a system can be
understood by considering classical stochastic fluctuations with
well defined probability distribution functions, for example,
photon shot noise in optical systems, the Casimir force between
plates, and the Lamb shift in atoms. However, it is not clear in
this context to what extent it is possible to describe additional
aspects of the dynamics of such quantum fluctuations using
classical stochastic concepts.

The dynamics of quantum fluctuations can be manifest when they couple
(weakly) to another external system. For example, when considering atoms
which interact with vacuum fluctuations of the electromagnetic field and
spontaneously emit photons, a process that can be described with high
precision by QED. Similarly, quantum fluctuations can affect spacetime
curvature. In this context, the possible gravitational collapse
of strong quantum fluctuations towards black holes was considered
recently\cite{eichler}, as well as issues concerning classicality
and decoherence of quantum fluctuations\cite{linde,polarski}
during and after inflation, and their time evolution
\cite{garriga,bucher}. Additionally, the possible dissipative
effects \cite{reynaud} of vacuum fluctuations on arbitrary
motions in the vacuum were considered.

We propose here a formalism to describe the dynamics of quantum
fluctuations in terms of classical stochastic processes which can
complement the standard representation in terms of time-dependent
quantum correlation functions. This formalism may enable better
understanding of the onset of classicality in quantum systems,
and in some cases may become a useful tool for performing
calculations, in particular numerical simulations of quantum systems.

We represent the vacuum of weakly
interacting quantum field theories in the Heisenberg picture as a linear
combination of randomly distributed  pseudoclassical paths (PCP's) that do
not interfere between themselves, and obey classical equations of
motion with constrained initial conditions.
Every physical observable is assigned a time-dependent value on
each PCP in a way that respects the uncertainty principle. In consequence,
some of the standard algebraic relations between quantum observables are
not necessarily fulfilled by their pseudoclassical values on paths. We
define collective observables which depend on a large number of
independent degrees of freedom, and show that their values on PCP's
approximately satisfy standard algebraic relations between classical
observables. We use the PCP's formalism to represent the dynamics of
quantum fluctuations of collective observables in terms of classical
stochastic processes with unconstrained initial conditions.

Our formalism can be applied to closed systems without
explicit reference to observers, measurement, or an environment.
We focus in this paper on fluctuations in the vacuum for
simplicity and concreteness, but our formalism can be generalized
in a straightforward way to describe quantum fluctuations in
other  states of the system, as we outline in section $V$.

Our formalism is somewhat reminiscent of the Feynmann path
integral formulation (FPI) \cite{Feynmann} and the consistent
histories formalism of Gell-Mann and Hartle \cite{GMH} in that it
involves paths as basic elements. However, in the FPI formalism,
the vacuum is considered as a superposition of coherent paths
that do not necessarily fulfill classical equations of motion and
produce interference patterns in Green's functions. In the
consistent histories approach all such coherent paths are grouped
into equivalence classes or non-detailed histories that are
approximately incoherent. In our formalism the paths obey
classical equations of motion and do not interfere at all, but
standard algebraic relations are not satisfied on them.

The notion that we propose for the emergence of classical
behaviour in the specific context that we discuss is quite
different than the case which is often discussed in the
literature. In the standard case which relies on the notion of
coherent states, for example, in cosmology, or in finite
temperature field theory, the emergence of classicality is a
consequence of the system being in a highly occupied state. If
the number of quanta in a given state is small, as in the vacuum
state of a quantum field theory in Minkowski space, there is no
notion of classicality. We show that some aspects of classical
behaviour can emerge also for states that are not highly
populated. A major difference between our formalism and the
coherent states formalism is that we identify a classical
behaviour associated to a certain class of observables of the
system, whereas in the formalism of coherent states classicality
is associated with a certain class of quantum states.

The paper is organized as follows, in section II we define the
concept of PseudoClassical Paths (PCP's), in section III we
investigate the properties of PCP's in quantum mechanics and free
quantum field theories, and determine the connection between
quantum expectation values and ensemble averages over PCP's. In
section IV we define collective observables and show that their
quantum fluctuations can be described classically in terms of
statistical averages over PCP's. In section V we outline the
extension of our formalism to weakly interacting quantum field
theories in states other than the vacuum.
In Section VI we compare our formalism to the formalism of
coherent states.
Section VII contains summary of results.

\section{PseudoClassical Paths}

Consider a quantum system with a given complete set
$\{Q_i,i=1,\dots,\hbox{\Large\em n}\}$ of \hbox{\Large\em n}
commuting observables. Each path corresponds to a set
$\{|q_i,i=1,\dots,\hbox{\Large\em n}\rangle\}$ of possible
expectation values in the vacuum for operators in $\{Q\}$. The
sets $\{|q_i\rangle\}$ span the configuration space of the quantum
system in its ground state $|0\rangle$. The probability of any
specific point $|q_i\rangle$ in configuration space to be
realized is given by $|\psi(q)|^2$, $\psi(q)$ being the
wavefunction of the vacuum in the $Q$-representation.

We assign at any point $|q_i\rangle$ the ``pseudoclassical"  value
$(Q_i)_{cl}=q_i$ to each operator in $\{Q\}$, and thus define on
configuration space \hbox{\Large\em n} independent random
variables $q_i$ whose statistics is determined by $|\psi(q)|^2$.
Furthermore, as we will show, the action on the vacuum of a large
class of observables $O$ (which may or may not commute with
observables in $\{Q\}$) can be expressed in a unique way as the
action of a polynomial function ${\cal P}_o$ of observables in
$\{Q\}$, $O|0\rangle={\cal P}_o(Q_i)|0\rangle$. Consequently we
may assign to each of such operators a random variable which is
the same polynomial function of the independent random variables
$q_i$ associated with operators in $\{Q\}$. At every point
$|q\rangle$ in configuration space we assign the
``pseudoclassical" value $O_{cl}(|q\rangle)={\cal P}_o(q_i)$ to
the observable $O$. In addition, in the Heisenberg picture, time
dependence of $O$ is given by the hermitian operator
$O(t)=e^{+iHt} O e^{-iHt}$, where $H$ is the hamiltonian of the
system. The observable $O(t)$ can be associated following the
same algorithm with a new random variable ${\cal P}_o(q_i,t)$
defined on configuration space. Thus, each point in configuration
space $|q_i\rangle$ is assigned a time-dependent
``pseudoclassical" value for each observable $O$ of the system,
that is a path whose realization probability is given by
$|\psi(q)|^2$. We will show that time evolution of the values
assigned on each of these paths to the observables in $\{Q\}$ and
their conjugate momenta obey classical equations of motion
with constrained initial conditions.

We call the paths ``pseudoclassical paths" (PCP's). They are not
classical paths, because the uncertainty principle prevents the
values assigned to observables which do not commute with
operators in $\{Q\}$ from satisfying standard algebraic relations
between classical observables:  it turns out that for such
observables $(O(t)_{cl})^{*} O(t)_{cl}$ is not necessarily equal
to $(O^{\dagger}(t) O(t))_{cl}$.

However, as we shall show, when $O$ depends on a large number of
degrees of freedom the differences $(O_{cl})^{*} O_{cl} -
(O^{\dagger} O)_{cl}$ are distributed on configuration space with
zero mean and negligible dispersion. In other words, classical
algebraic relations  are approximately recovered on almost all
paths for such observables. We call them ``collective
observables". We note that quantum fluctuations of collective
observables behave classically, and we may borrow classical
concepts to define their typical lifetimes. According to the
description we have outlined, we may attribute classical aspects
to quantum fluctuations of collective observables, in particular,
decoherence.

\section{PseudoClassical Paths  in Quantum mechanics
and in
 Quantum Field Theory}

To discuss our proposal in a concrete and tractable setup we
consider a free complex relativistic scalar field $\Phi(x)$ in
Minkowski space-time whose lagrangian density is given by
\footnote{we use units in which $\hbar=1$, $c=1$.}
\begin{equation}
{\cal L} = \partial_{\mu}\Phi^* \partial^{\mu}\Phi  - M^2 \Phi^*
\Phi.
\end{equation}
The vacuum state of this free theory provides the simplest
example for which the ideas we have introduced above can be
developed and, in addition, serves as a zero'th order perturbative
approximation to the actual vacuum of any weakly interacting
theory. Later on in section V we will outline how to extend our description to
higher orders of perturbation theory, and to states other than the vacuum.

The hamiltonian of the system
\begin{equation}
{\cal H} = \int d^{3}{\vec x} \left(\Pi^*({\vec x}) \Pi({\vec x}) +
\partial_{j}\Phi^*({\vec x}) \partial_{j}\Phi({\vec x}) +
M^2 \Phi^* \Phi \right),
\end{equation}
can be expanded in terms of Fourier modes of the field
$\Phi({\vec x})$ and its conjugate momentum $\Pi^*({\vec x})$:
\begin{equation}
\label{field}
\Phi({\vec x}) = \sum_{n_1,n_2,n_3=-N}^{+N} \frac{1}{X^{3/2}}
q_{n_1,n_2,n_3}
exp\left(2\pi i (\sum_{j=1}^{3} n_j x_j)/X \right)
\end{equation}
\begin{equation}
\label{momentum}
\Pi^{*}({\vec x}) = \sum_{n_1,n_2,n_3=-N}^{+N} \frac{1}{X^{3/2}}
p^*_{n_1,n_2,n_3}
exp\left(-2\pi i (\sum_{j=1}^{3} n_j x_j)/X \right),
\end{equation}
in order to express ${\cal H}$ in terms of decoupled harmonic
oscillators,
\begin{equation}
{\cal H} = \sum_{\vec n}
\left( p^*_{\vec n}  p_{\vec n} + \kappa^2_{\vec n}
q^*_{\vec n}  q_{\vec n} \right),
\end{equation}
where $\kappa_{\vec n}=\left[(\frac{2\pi}{X})^2 |{\vec n}|^2 +
M^2\right]^{1/2}$, and $x_j$ are cartesian coordinates.  We impose
periodic boundary conditions on the three dimensional box
$\left[0,X\right] \times \left[0,X\right] \times
\left[0,X\right]$. The size of the box $X$ serves as an infrared
cutoff, and $x_{min} = X/N$ serves as an ultraviolet cutoff.

We introduce normalized canonical operators $(Q,P)_{\vec n}$, $
Q_{\vec n}= {\sqrt{\kappa_{\vec n}}} q_{\vec n} $, and $P_{\vec
n}= \frac{1}{\sqrt{\kappa_{\vec n}}} p_{\vec n}$, which obey
canonical commutation relations: $[Q_{\vec n},P^{\dagger}_{\vec
m}]=[Q^{\dagger}_{\vec n},P_{\vec m}]=i \delta_{{\vec n}{\vec
m}}$ while any other commutators between them vanish. The
hamiltonian can be expressed in terms of the $P$'s and $Q$'s, $
{\cal H} = \sum_{\vec n} \kappa_{\vec n} \left(P^{\dagger}_{\vec
n}
 P_{\vec n} + Q^{\dagger}_{\vec n}  Q_{\vec n} \right).
$

Let us focus on the Hilbert space of states of one of the modes,
labeled by some generic $\vec n$. The hamiltonian for this
specific mode is $H=\kappa \left(P^{\dagger} P + Q^{\dagger} Q
\right)$ (where $\kappa=\kappa_{\vec n}$), and we denote its
vacuum state by $|0\rangle$. We now show that, for operators $O$
that can be expressed as a polynomial in $Q$, $P$ and their
hermitian conjugates   $O\equiv{\cal P}
(Q^{\dagger},Q,P^{\dagger},P)$, it is possible to express
$O|0\rangle$ only in terms of a polynomial ${\cal P}_o$ of $Q$
and its hermitian conjugate $Q^{\dagger}$, $O|0\rangle= {\cal
P}_o(Q^{\dagger},Q)|0\rangle$.

First we would like to determine how to express the action of $P$
and $P^\dagger$ on the vacuum. We note that the operators $(Q,P)$
can be written in terms of hermitian components
$Q=\frac{1}{\sqrt{2}} \left(Q_R + i Q_I \right)$ and
$P=\frac{1}{\sqrt{2}} \left(P_R + i P_I \right)$.  The hermitian
operators $Q_R,P_R,Q_I,P_I$ obey canonical commutation relations
$[Q_R,P_R]=[Q_I,P_I]=i $, and any other commutators between them
vanish. Therefore, $a_R|0>=\frac{1}{\sqrt{2}}(Q_R+iP_R)|0>=0$ and
$a_I|0>=\frac{1}{\sqrt{2}}(Q_I+iP_I)|0>=0$, so that
$Q_R|0\rangle=-i P_R|0\rangle$, and $Q_I|0\rangle=-i
P_I|0\rangle$. We now substitute into these two equalities the
relations $Q_R= \frac{1}{\sqrt{2}}(Q+Q^{\dagger})$, $P_R=
\frac{1}{\sqrt{2}}(P+P^{\dagger})$, $Q_I=-i
\frac{1}{\sqrt{2}}(Q-Q^{\dagger})$, and $P_I=-i
\frac{1}{\sqrt{2}}(P-P^{\dagger})$. Simple algebra leads to the
key relations,
\begin{eqnarray}
 \label{keypoint1}
P|0\rangle&=&iQ|0\rangle \\
P^{\dagger}|0\rangle&=&iQ^{\dagger}|0\rangle.
 \label{keypoint2}
\end{eqnarray}
Note that eq.(\ref{keypoint1}) is not the conjugate of
eq.(\ref{keypoint2}).

To proceed with the proof of our claim we look at terms of the
form $PQ$. From the commutation relations of $P$ and $Q$ we obtain
\begin{eqnarray}
 \label{relat}
P^\dagger Q|0\rangle&=&(QP^{\dagger}-i)|0\rangle \\
PQ|0\rangle &=& QP|0\rangle,
 \label{relat1}
\end{eqnarray}
and the hermitian conjugates of both relations. From
(\ref{keypoint2}) we conclude that
\begin{equation}
 \label{keypoint3}
 QP^{\dagger}|0\rangle=iQ^{\dagger}Q|0\rangle,
\end{equation}
which upon substitution into eq.(\ref{relat}) yields
\begin{equation}
P^{\dagger}Q|0\rangle=i(Q^{\dagger}Q-1)|0\rangle.
 \label{keypoint4}
\end{equation}
Substituting eq.(\ref{keypoint1}) into eq.(\ref{relat1}) we obtain
\begin{equation}
PQ|0\rangle=iQ^2|0\rangle.
 \label{keypoint41}
\end{equation}
Additionally, since $ H|0\rangle=\kappa |0\rangle$, $P^\dagger
P+Q^\dagger Q=1$, so we conclude that
\begin{equation}
(P^{\dagger} P)|0\rangle=(1-Q^{\dagger} Q)|0\rangle,
 \label{keypoint5}
\end{equation}
which is, of course, the same result as we would have obtained by
using eq.(\ref{keypoint1}) and eq.(\ref{keypoint4}) in sequence.

Using eqs.(\ref{keypoint1}-\ref{keypoint5}) it is possible to
express the action on the vacuum of any monomial (and therefore
any polynomial) with arbitrary powers and order of $Q$,
$Q^\dagger$, $P$ and $P^\dagger$ as a polynomial in $Q$,
$Q^\dagger$. For example, commute all the $P$'s to the right most
position so they act directly on $|0\rangle$ and use
eq.(\ref{keypoint1}) to express them in terms of $Q$. Then
commute all the $P^\dagger$'s to the right most position and use
eq.(\ref{keypoint2}) to express them in terms of $Q^\dagger$.

This procedure is guaranteed to be well defined, that is, produce
a unique result for all monomials that are related to each other
by permuting commuting operators. To see this, assume that two
such monomials lead to the same polynomial in $Q$, $Q^\dagger$.
Then they should be identical when acting on the vacuum, so,
their difference (which can also be expressed as a monomial in
$Q$, $Q^\dagger$, $P$ and $P^\dagger$) vanishes when acting on the
vacuum. Conversely, assuming that the same monomial leads to two
different polynomials, and using the same logic we conclude that
the difference between the two polynomials (which can also be
expressed a polynomial in $Q^{\dagger},Q$) must be the polynomial
zero, which means that the resulting polynomials are identical.

The operators $Q^{\dagger},Q$ form a complete representation on
the Hilbert space of states that we are considering. In this
representation each point $|q\rangle$ in configuration space is labeled
by a single complex number $q$ and the wavefunction of its vacuum
state is
\begin{equation}
\psi(q)=\frac{\sqrt{2}}{\sqrt{\pi}}e^{-|q|^2},
 \label{gaussian}
\end{equation}
normalized such that
\begin{equation}
 \label{gaussian1}
 \langle0|0\rangle=\int dq
|\psi(q)|^2=\frac{2}{\pi} \int dq e^{-2 |q|^2}=1.
\end{equation}

Quantum fluctuations of $Q^{\dagger},Q$ can be thought of as
random events that can be described by complex conjugates random
variables $q^{*},q$ defined on configuration space, whose
distribution function is the gaussian $|\psi(q)|^2$. Note that we
have chosen the same notation for a point $|q\rangle$ in
configuration space and for the random variable $q$ whose value
at each $|q\rangle$ is $Q_{cl}=q$. Since $O|0>={\cal
P}_o(Q^{\dagger},Q)|0\rangle$, we assign to each point
$|q\rangle$ the value $O_{cl}={\cal P}_o(q^{*},q)\equiv o(q^*,q)$
for the observable $O$.  For example, according to
eq.(\ref{keypoint1}) we assign the variable $p=iq$ to the
observable $P$,  according to eq.(\ref{keypoint2}) we assign
$p^{\dagger}=iq^{*}$ to $P^{\dagger}$, $q2=q^{*}q$ to
$Q^{\dagger} Q$, and according to eq.(\ref{keypoint5})
$p2=(1-q^{*}q)$ to the observable $P^{\dagger} P$ and $h=\kappa$
to the hamiltonian $H$.

As we have pointed out previously, in the Heisenberg picture the
observable $O$ at time $t$ is given by $O(t)=e^{+iHt} O e^{-iHt}$
which gets associated according to the rules outlined above
with a new random variable $o(t)$. For example,
$Q(t)|0\rangle=e^{i\kappa t} Q|0\rangle$ and $P(t)=i e^{i\kappa
t} Q|0\rangle$, and therefore for each $|q\rangle$ we assign to
these operators the values $q(t)=e^{i\kappa t} q$ and $p(t)=i
e^{i\kappa t} q = i q(t)$. Of course, $H(t)=H$ and therefore
$h(t)=h$ at any given time. In this sense each point $|q\rangle$
in configuration space is associated in the Heisenberg picture
with a path, since any physical observable can be assigned a
time-dependent ``pseudoclassical" value on each of them. The
paths are random since their realization is determined
statistically, but when realized they are in some sense
deterministic and obey classical equations, as we will show
later. The vacuum state $|0\rangle=\int dq \psi(q) |q\rangle$ can
then be thought of as a linear combination of such paths. The
probability of each path to be realized due to quantum
fluctuations is given at any time by the gaussian distribution
$|\psi(q)|^2$.

As advertised, the ``pseudoclassical" values assigned to $Q(t)$,
$Q_{cl}(t)=q(t)$, and its conjugate momentum $P(t)$ obey the
appropriate classical equations of motion for the harmonic
oscillator \footnote{Similarly, we could have started from
relations $\langle 0|a_R^{\dagger}=0$ and $\langle
0|a_I^{\dagger}=0$ to obtain $\langle 0|P^{\dagger}= -i\langle
0|Q^{\dagger}$ and $\langle 0|P=-i\langle 0|Q$ instead of eqs.
(\ref{keypoint1}),(\ref{keypoint2}). Following the same rules we
have stated above we obtain the other set of possible solutions
to the classical equations $q(t)=exp(-i\kappa t) q$, $p(t)=-i
exp(-i\kappa t) q$.}, $dp(t)/dt=-\partial H/\partial q^{*}$ and
$dq(t)/dt=\partial H/\partial p^{*}$, with constrained initial conditions
which have to be consistent with the uncertainty principle,
and in particular with eqs.(\ref{keypoint1}), (\ref{keypoint2}).
The solutions are periodic
functions whose period is $T=\frac{2\pi}{\kappa}$, and we may
identify the characteristic time scale of these paths with
$T$. When considering more general operators that are generic
polynomials of $q(t)$ it is possible to construct solutions that
have different characteristic time scales, in complete analogy
with classical Fourier analysis relating time- and
frequency-dependent functions.

We now proceed to show that the quantum vacuum expectation value
of an operator $O$ is equal to the ensemble average of its
corresponding random variable $o$ over the ensemble of PCP's,
$\langle 0|O|0\rangle=\langle o \rangle_{PCP}\equiv\langle O_{cl}
\rangle_{PCP}$.  By our construction we can express the vacuum
expectation value  of a general polynomial only in terms of a
polynomial in $Q$ and $Q^\dagger$, $\langle 0|O|0 \rangle =\langle
0|{\cal P}_o(Q^{\dagger},Q)|0\rangle$, which is given by
\begin{equation}
\langle0|{\cal P}_o(Q^{\dagger},Q)|0\rangle=\int dq \frac{2}{\pi}
{\cal P}_o(q^{*},q) e^{-2|q|^2}.
\end{equation}
By definition  $\int dq \frac{2}{\pi} {\cal P}_o(q^{*},q)
e^{-2|q|^2}=\langle o(q^*,q)\rangle_{PCP}\equiv\langle
O_{cl}\rangle_{PCP}$. The PCP's are incoherent: they do not
interfere when we compute ``pseudoclassical" average values.

Similarly,  expressions can be obtained for the second moments of
random variables. First,  denote $\langle
o2\rangle_{PCP}\equiv\langle (O^{\dagger} O)_{cl}\rangle_{PCP}$.
By construction $O|0\rangle={\cal P}_o(Q^{\dagger},Q)|0\rangle$,
which implies that $ \langle 0|O^{\dagger} O|0\rangle = \langle
0|({\cal P}_o(Q^{\dagger},Q))^{\dagger} {\cal
P}_o(Q^{\dagger},Q)|0\rangle$. Now notice that $\langle 0|{\cal
P}_o(Q^{\dagger},Q)^{\dagger} {\cal
P}_o(Q^{\dagger},Q)|0\rangle=\langle (O_{cl})^{*}
O_{cl}\rangle_{PCP}$ which by definition satisfies $\langle
(O_{cl})^{*} O_{cl}\rangle_{PCP}=\langle o^* o\rangle$. Thus we
have shown that
\begin{equation}
\label{second} \langle o^* o\rangle_{PCP}-\langle o2\rangle_{PCP}=
\langle (O_{cl})^{*}  (O_{cl})-(O^{\dagger} O)_{cl}\rangle=0.
\end{equation}

Equation (\ref{second}) holds on average over the ensemble of
PCP's, but it does not necessarily hold on each one of the paths
nor even on any of them. The ``pseudoclassical" values assigned
on a path to physical observables which do not commute with
$Q^{\dagger},Q$ do not obey standard algebraic relations between
classical observables as a consequence of the uncertainty
principle, that is, for such operators $(O_{cl})^{*} (O)_{cl} -
(O^{\dagger} O)_{cl}$ does not necessarily vanish on each path.
This is  why we call such paths pseudoclassical, even though they
obey classical equations of motion. As a simple example let us
consider the operator $P$. We know that $(P_{cl})^{*}
(P_{cl})=(iQ_{cl})^{*} (iQ_{cl})=(Q^{\dagger}Q)_{cl}
=1-(P^{\dagger}P)_{cl}$, so
$(P_{cl})^{*}(P_{cl})-(P^{\dagger}P)_{cl}=
1-2(P^{\dagger}P)_{cl}$ which clearly does not vanish identically
on a given PCP, but only on the average because $\langle
0|P^{\dagger}P|0\rangle=1/2$.

\section{Collective observables and
Classical dynamics of their quantum fluctuations}

Let us now return to the full field theoretic setup and consider
generic local Lorentz-covariant field operators of the type
$O({\vec x})={\cal P}(\Phi^{\dagger}({\vec x}),\Phi({\vec
x}),\Pi^{\dagger}({\vec x}),\Pi({\vec x}))$, which involve
``collective" dynamics of all (or many) of the single harmonic
modes in the box. Recall that we have regularized the theory in
the UV and in the IR, so the number of modes is finite. The free
field theoretic vacuum state is therefore a finite tensor product
of the vacuum states of each one of the single decoupled
oscillators. It can  be expressed as a linear combination of
pseudoclassical field configurations (PCFC's) which are tensor
products of the PCP's of single harmonic oscillator modes, and
therefore evolve in time according to classical field equations,
since each of the PCP's evolve in time according to its own
classical equations of motion. Every field theoretic observable
can be expressed as a function of single mode observables and thus
assigned a time-dependent ``pseudoclassical" value on each field
configuration.

Some field theoretic observables have the property that
$(O_{cl})^{*} O_{cl} - (O^{\dagger} O)_{cl} \sim 0$, not only in
the average sense of eq.(\ref{second}) but in the stronger sense
that the dispersion over the ensemble of PCFC's of this
variable  is much smaller than the mean of the variable
$(O_{cl})^{*} (O)_{cl}$:
\begin{equation}
\label{condition} \sigma_O=\sqrt{\left\langle\left|(O_{cl})^{*}
(O_{cl})\!-\!(O^{\dagger} O)_{cl}\right|^2\right\rangle_{PCP}}~
\ll ~ \left\langle\left|O_{cl}\right|^2\right\rangle_{PCP}.
\end{equation}
That is, standard classical algebraic relations are approximately
recovered on most of the PCFC's for such observables. Condition
(\ref{condition}) is naturally satisfied for observables $O$
which involve many independent single mode degrees of freedom. We
call them ``collective observables". According to the central
limit theorem, in this case the probability distribution of the
variable $o^* o - o2$ is gaussian and therefore its higher
statistical moments are also suppressed. Condition (\ref{condition})
effectively removes the constraints on the allowed initial conditions
for the classical values of collective operators, and therefore
on the initial conditions of the corresponding classical
stochastic processes.

Condition (\ref{condition}) does not necessarily hold for
observables defined on the Hilbert space of states of a single
harmonic mode. For example, let us reexamine the operator
$(P_{cl})^{*} (P_{cl})-(P^{\dagger} P)_{cl}=(i Q_{cl})^* (i
Q_{cl})-(1-(Q^{\dagger} Q)_{cl})=-1 + 2 (Q_{cl})^{*} Q_{cl}$. The
mean value of this random variable on configuration space, and
$\sigma_P$, can be computed directly from the distribution
function (\ref{gaussian}) of variables $Q_{cl},(Q_{cl})^{*}$. The
mean value vanishes according to (\ref{second}), but
$\sigma_P=1=2 <|Q_{cl}|^2>=2 <|P_{cl}|^2>$ is not negligible.

For collective observables that can be
expressed as a direct sum of many independent single mode operators,
\begin{equation}
 \label{operator}
O({\vec x}) = \sum_{n_1,n_2,n_3=-{N}}^{+ {N}} \frac{1}{X^{3/2}}
O_{n_1,n_2,n_3} exp\left(-2\pi i (\sum_{j=1}^{3} n_j x_j)/X
\right),
\end{equation}
such that $O_{\vec n} O_{\vec m}|0\rangle=O_{\vec n}|0\rangle
\bigotimes O_{\vec m}|0\rangle$ if ${\vec n} \not= {\vec m}$, we
find that the random variable $o^* o - o2$ is a quadratic form of
${\cal N}$ independent random variables, ${\cal N}$ being the
effective number of contributing single modes, that is the number
of coefficients $O_{n_1,n_2,n_3}$ which are substantially
different from zero. Thus, the dispersion $\sigma_O$ is suppressed
by a factor $1/\sqrt{{\cal N}}$ relative to the mean squared value
of the variable $o^* o$: $\sigma_O \sim \frac{1}{\sqrt{{\cal N}}}
\langle|O_{cl}|^2\rangle_{PCP}$. We may attribute
classical properties to the dynamics of quantum fluctuations of such
collective observables.

Let consider, for example, the operator $\Pi({\vec x})$ introduced
in eq.(\ref{momentum})
\begin{equation}
\Pi({\vec x}) = \sum_{n_1,n_2,n_3=-N}^{+N} \frac{1}{X^{3/2}}
{\sqrt{\kappa_{\vec{n}}}} P_{\vec{n}} exp\left(-2\pi i
(\sum_{j=1}^{3} n_j x_j)/X \right).
\end{equation}
Following our rules we assign to $\Pi({\vec x})$ the random
variable
\begin{equation}
\pi({\vec x}) = \sum_{n_1,n_2,n_3=-N}^{+N} \frac{1}{X^{3/2}}
{\sqrt{\kappa_{\vec{n}}}} (i q_{\vec{n}}) exp\left(-2\pi i
(\sum_{j=1}^{3} n_j x_j)/X \right),
\end{equation}
defined in terms of $N^3$ independent random variables $q_{\vec
n}$. Therefore,
\begin{equation}
(\pi({\vec x}))^* \pi({\vec x}) = \sum_{\vec n,\vec m}
\frac{1}{X^{3}}
{\sqrt{\kappa_{\vec n} \kappa_{\vec m}}} (q^*_{\vec n} q_{\vec m})
exp\left(2\pi i ({\vec n}-{\vec m})\cdot {\vec x}/X \right).
\end{equation}
On the other hand,
\begin{equation}
(\Pi({\vec x}))^{\dagger} \Pi({\vec x}) = \sum_{\vec n,\vec m}
\frac{1}{X^{3}}
{\sqrt{\kappa_{\vec n} \kappa_{\vec m}}} (P^{\dagger}_{\vec n} P_{\vec m})
exp\left(2\pi i ({\vec n}-{\vec m})\cdot {\vec x}/X \right),
\end{equation}
so that, using $(P^{\dagger}_{\vec n} P_{\vec
m})_{cl}=\delta_{\vec n\vec m}- (Q^{\dagger}_{\vec n} Q_{\vec
m})_{cl}$, we find that the corresponding random variable is
\begin{equation}
\pi2({\vec x})= \sum_{\vec n,\vec m} \frac{1}{X^{3}}
{\sqrt{\kappa_{\vec n} \kappa_{\vec
m}}}\left[\delta_{\vec{m}\vec{n}}-q^*_{\vec n} q_{\vec m}\right]
exp\left(2\pi i ({\vec n}-{\vec m})\cdot {\vec x}/X \right),
\end{equation}
and therefore
\begin{equation}
(\pi({\vec x}))^* \pi({\vec x})-\pi2({\vec x})=-\sum_{\vec n}
\frac{1}{X^{3}} \kappa_{\vec n} + 2 \sum_{\vec n,\vec m}
\frac{1}{X^{3}} {\sqrt{\kappa_{\vec n} \kappa_{\vec m}}} (p^*_{\vec n}
p_{\vec m}) exp\left(2\pi i ({\vec n}-{\vec m})\cdot {\vec x}/X \right),
 \label{pisquare}
\end{equation}
which, as anticipated, is a quadratic form in many independent
gaussian random variables. Its mean is zero as expected because
$\langle (p^*_{\vec n} p_{\vec
m}\rangle_{PCP}=\frac{1}{2}\delta_{\vec n\vec m}$ from
(\ref{gaussian}). The dominant contribution to the variables
appearing in eq.(\ref{pisquare}) comes from  modes whose momentum
is about the ultraviolet cutoff, so that effectively ${\cal N}
\sim N^2$. The dispersion $\sigma_{\Pi}$ is of order $\sqrt{\cal
N} \sim N$. On the other hand the average value of $(\pi({\vec
x}))^* \pi({\vec x})$ is of order ${\cal N} \sim N^2$. We see
explicitly that $\Pi(\vec{x})$ is a collective observable.

As another concrete illustration of our formalism and for a more
detailed discussion of lifetimes of quantum fluctuations of
collective observables let us consider the collective operator
(introduced in \cite{eichler}),
\begin{eqnarray}
&{\cal H}_V =& \frac{1}{2 X^3} \sum_{{\vec n},{\vec m} \neq 0}
\frac{1}{\sqrt{\kappa_{\vec n} \kappa_{\vec m}}} F_V({\vec
n}-{\vec m},X) \left[ \kappa_{\vec n} \kappa_{\vec m}
P^{\dagger}_{\vec n} P_{\vec m} + \kappa_{{\vec n}\cdot{\vec m}}
Q^{\dagger}_{\vec n}  Q_{\vec m} \right] + {\it h.c.},
\end{eqnarray}
which describes the energy contained in a finite volume V
enclosed in the 3D box $[0,X]^3$. On the right hand side of the
expression we have introduced the notation $\kappa_{{\vec
n}\cdot{\vec m}}=(\frac{2\pi}{X})^2 ({\vec n} \cdot {\vec m}) +
M^2$, where the $M$ is the mass of the scalar field. The function
$F_V({\vec k},X)=\int_{V} d^3{\vec x} e^{-2\pi i\frac{{\vec
k}\cdot{\vec x}}{X}}$ can be evaluated analytically for simple
geometries, for example, a box with dimensions $L_1,L_2,L_3 < X$.
In this particular case: $F_V({\vec k},X)=V$, if ${\vec k}=0$;
and $F_V({\vec k},X)=\prod_{{\it j}=1,2,3} \frac{i X}{2 k_{\it j}
\pi}(e^{2\pi i k_{\it j} L_{\it j}/X}-1)$, if ${\vec k}\neq 0$.

As noticed in \cite{eichler}, if $V$ is the whole box then the
vacuum $|0\rangle$ of the system is an eigenstate of $H_V$, so it
does not fluctuate. For partial volumes whose typical size
$V^{1/3}$ is parametrically larger than the ultraviolet cutoff
$X/N$,  the vacuum $|0\rangle$ is no longer an eigenstate of
$H_V$, and therefore $H_V$ fluctuates quantum mechanically.

Using our rules it is straightforward to obtain the random
variable $h_V$ assigned to $H_V$,
\begin{equation}
{\it h}_V(t) = {\it h}^0_V + \frac{1}{X^3} \sum_{{\vec n},{\vec
m}\neq 0} {\it Re}\left[ \frac{F_V({\vec n}-{\vec
m},X)}{\sqrt{\kappa_{\vec n} \kappa_{\vec m}}}
\left(\kappa_{{\vec n}\cdot{\vec m}} - \kappa_{\vec n}
\kappa_{\vec m} \right) q^*_{\vec n} q_{\vec m}
\right]exp(i(\kappa_{\vec n}+\kappa_{\vec m})t),
 \label{hv}
\end{equation}
where ${\it h}^0_V = \frac{V}{X^3} \sum_{\vec n} \kappa_{\vec n}$
is the average value of quantum energy fluctuations in the volume
$V$ and $q_{\vec n}, q_{\vec m}$ are independent gaussian random
variables.

We follow \cite{landau} to define the lifetime of classical
stochastic fluctuations: first, we construct the time-dependent
correlation function $f(t_1-t_2)={\it Re}\left((O_{cl}(t_1))^*
O_{cl}(t_2)\right)$; then, we define the lifetime of the
fluctuation as the inverse of the width in frequency $\omega$ of
the Fourier transform $F(\omega)$. The ensemble average $\langle
f(t_1-t_2)\rangle_{PCP}= \langle O^{\dagger}(t_1) O(t_2) +
O^{\dagger}(t_2) O(t_1)\rangle$ coincides with the standard
definition of lifetime of quantum fluctuations.
Our definition of the 'lifetime' of the fluctuations is the
definition that is usually used to describe the time scale of a
virtual excitation and is not related to the more common
definition of transition time from a quantum state to other
quantum state induced by a coupling term in the hamiltonian. We
believe that our definition is the appropriate definition to
describe fluctuations of a quantum system without any reference
to an environment or an external perturbing system.

Notice that $\omega$ is a scalar, for example, in the expansion
(\ref{hv}) above $\omega=\kappa_{\vec n}+\kappa_{\vec m}$, so
that all the different modes labeled by vectors ${\vec n}$ and
${\vec m}$ which give the same $\omega$ contribute to a unique
Fourier mode. The integration over the angles of these vectors
contributes to the coefficient of this mode, but not to the width
which goes into the definition of the lifetime. Modes that have
the same frequency  oscillate coherently, and according to the
above definition, their lifetime would be infinite. Finite
lifetime result from the superposition of many modes with
different frequencies. The width in $\omega$ gives the range over
which these frequencies are distributed.

The main contribution to $h_V(t)$ in eq.(\ref{hv}) comes from
modes whose momentum is about the ultraviolet cutoff
$|\vec{n}|\sim|\vec{m}|\sim N$. Its Fourier transform is peaked
around $\omega_{UV}$, and the dispersion in $\omega$, $\Delta\omega$,
is of order $\omega_{UV}$. Therefore, we find that lifetime of
dominant contributions to the energy fluctuation
$t\simeq\frac{2\pi}{\Delta\omega}$ is inversely proportional to
the ultraviolet cutoff of the theory. Subdominant contributions
from modes $|\vec{n}|\sim|\vec{m}| < N$ have a longer lifetime, as
noticed in \cite{eichler}.

\section{Extensions}

\subsection{Extension to states other than the vacuum}

The description of the free vacuum in terms of PCFC's that we have
presented in the previous sections can be generalized to describe
other states of the free theory, stationary or non-stationary.
In this subsection we show how to carry out this
extension. In the next subsection we will show how the description
in terms of PCFC's can be further generalized to quantum states
in weakly interacting theories.

Since any state of a free quantum field theory
can be represented in a basis of tensor products of single mode
stationary states, we will first show how to extend the formalism
of PCP's to stationary excited states of a single harmonic
mode. Our immediate goal is to
express the action of each single mode operator $O$ on a generic
single mode stationary state $|\psi \rangle$ in terms of some specific
function $F^{\psi}_o$ (which is not necessarily a polynomial) of $Q$ and
$Q^{\dagger}$ on $|\psi \rangle$: $O|\psi
\rangle=F^{\psi}_o(Q^{\dagger},Q)|\psi \rangle$.

Recall that $Q$ can be written in terms of two hermitian
components $Q=\frac{1}{\sqrt{2}} \left(Q_R + i Q_I \right)$ and,
therefore, the stationary states of the single harmonic mode that
we are considering  are labeled by two natural numbers $|J1,J2
\rangle = \frac{1}{\sqrt{J1!}} \frac{1}{\sqrt{J2!}}
(a^{\dagger}_R)^{J1} (a^{\dagger}_I)^{J2} |0 \rangle$,
$J1,J2=0,1,2,...$, where we are using the notation introduced in
section III. For simplicity we will restrict this discussion  to
the sector of the Hilbert space generated by $a^{\dagger}_R$:
$|J1 \rangle = \frac{1}{\sqrt{J1!}} (a^{\dagger}_R)^{J1} |0
\rangle$.

The state $|J1 \rangle$  can be expressed in terms of the
Hermite polynomial $H_{J1}(Q_R)$ acting upon the vacuum,
\begin{equation}
\label{J1}
|J1 \rangle = \sqrt{\frac{1}{\sqrt{\pi} 2^{J1} J1!}} H_{J1}(Q_R)|0\rangle.
\end{equation}
The wavefunction of $|J1 \rangle$ vanishes at the zeroes of the
Hermite polynomial and therefore it is possible to invert
eq.(\ref{J1})
\begin{equation}
\label{J2} |0 \rangle = \sqrt{\sqrt{\pi} 2^{J1} J1!}
\left(H_{J1}(Q_R)\right)^{-1} |J1 \rangle,
\end{equation}
which allows us to express the vacuum as a function of the operator $Q_R$
acting upon the excited state $|J1 \rangle$.

We use eq.(\ref{J1}) to express $O|J1 \rangle =
\sqrt{\frac{1}{\sqrt{\pi} 2^{J1} J1!}} O H_{J1}(Q_R)|0 \rangle$ as
an operator acting on the free vacuum. Then, we use the rules we
stated in section III  to determine the action of this
operator on the vacuum as a polynomial in $Q_R$: $O H_{J1}(Q_R)|0
\rangle = P_{OH_{J1}}(Q_R)|0 \rangle$. Finally, we use
eq.(\ref{J2}) to express the free vacuum as a function of $Q_R$
acting on $|J1 \rangle$. This completes the algorithm that gives
the required expression $O|J1 \rangle = F^{J1}_o(Q_R)|J1
\rangle$. As each single step in the algorithm is uniquely
defined the resulting function is also uniquely defined.

As an example, let us apply the proposed algorithm to the
particular case of the momentum operator $P_R$ acting upon the
first excited harmonic mode $|1 \rangle$. First, express the
excited state by the first Hermite polynomial acting on the free
vacuum:
\begin{equation}
\label{excited}
|1 \rangle = \sqrt{\frac{1}{\sqrt{\pi} 2}} 2 Q_R|0 \rangle,
\end{equation}
to get
 \begin{equation}
\label{excited1.5}
 P_R|1 \rangle = \sqrt{\frac{1}{\sqrt{\pi} 2}} 2 P_R Q_R|0 \rangle.
 \end{equation}
Then, follow the standard rules stated in section III $P_R Q_R|0
\rangle = i(Q_R^2-1)|0 \rangle$, so that $P_R|1 \rangle =
\sqrt{\frac{1}{\sqrt{\pi} 2}} 2i (Q_R^2-1)|0 \rangle$. Finally,
invert eq.(\ref{excited})
\begin{equation}
\label{excited2}
P_R|1 \rangle = i \frac{Q_R^2-1}{Q_R}|1 \rangle.
\end{equation}

The algorithm can be applied in similar ways to time dependent
operators $O(t)$, so that we can apply the formalism of
pseudoclassical paths. Each PCP corresponds to a point $|q_r
\rangle$ in configuration space. The probability of such path to
be realized is now given by the modulus squared of the wavefunction of
the first excited state $|\psi_1(q_r)|^2$ at such point. Let us
point out again that the operator $1/Q_R$ is well defined on $|1
\rangle$ because $\psi_1(q_r)$ vanishes at $q_r=0$.

It is also possible generalize our formalism to non-stationary
quantum states of single harmonic modes. Any state $|\psi\rangle$
can be expanded in a basis of stationary states $|\psi
\rangle=\sum c_{J1} |J1 \rangle$. All finite linear combinations
of basis states can also be expressed as a polynomial in $Q_R$
acting upon the free vacuum $|\psi \rangle=\sum c_{J1} |J1
\rangle=\sum c_{J1} \sqrt{\frac{1}{\sqrt{\pi} 2^{J1} J1!}}
H_{J1}(Q_R)|0 \rangle= H(Q_R)|0 \rangle$. The subsequent steps of
the algorithm are identical to those defined for the case of a
single stationary state, in particular inversion is possible
because $H(Q_R)$ is a polynomial and the wave function of the
state vanishes at points where $H(Q_R)$ vanishes.

We would like to emphasize that the essential reason that
pseudoclassical values $p(t)$ and $q(t)$ assigned to observable
$P(t)$ and $Q(t)$ obey classical equations of motion on each PCP
for general quantum states is the operator equations that $Q_R(t)$
and $P_R(t)$ obey,
\begin{eqnarray}
\label{operators}
 \frac{dP_R(t)}{dt}&=&-Q_R(t) \nonumber \\
 \frac{dQ_R(t)}{dt}&=& P_R(t).
\end{eqnarray}
The operator equations (\ref{operators}) assure that the classical
equations of motion are valid when the operators act on any
quantum state
\begin{eqnarray}
\label{operators1}
 \frac{dP_R(t)|\psi\rangle}{dt} &=&
\frac{dF^{|\psi\rangle}_{P_R}(Q_R,t)|\psi\rangle}{dt} =
-Q_R(t)|\psi\rangle
\nonumber \\
 \frac{dQ_R(t)|\psi\rangle}{dt} &=& P_R(t)|\psi\rangle
= F^{|\psi\rangle}_{P_R}(Q_R,t)|\psi\rangle.
\end{eqnarray}
However, the particular relation between the
classical initial value of the generalized coordinate and its
conjugate momentum are different for each different quantum
state, as can be realized for example by comparing equations
(\ref{keypoint1}) and (\ref{keypoint2}) with equation
(\ref{excited2}). In other words, different sets of initial
conditions correspond to PCP's of different quantum states.

The description in terms of PCFC's can now be extended to quantum
field theoretic states other than the vacuum.
States that can be obtained as
a tensor product of single mode states can be described in terms of
PCFC's which are tensor products of the PCP's of single harmonic
oscillator modes. Other quantum field theoretic states that can be
obtained as a linear combination of tensor products can be written
in terms of a polynomial in $Q_{\vec n}$ and $Q^{\dagger}_{\vec m}$
acting upon the free field vacuum $|\psi \rangle = H(Q_{\vec
n},Q^{\dagger}_{\vec m})|0 \rangle$. The rest of the discussion is
identical to that for a single mode state.

\subsection{Extension to Interacting Quantum Field Theories}

In this subsection we outline how to extend the formalism of
PCFC's to the case in which the hamiltonian contains weak
interaction terms. This is not intended to be an exhaustive
study, rather we wish to show that there are no obvious problems
in the implementation of our formalism to interacting theories,
and thus show that it is not based on some particular properties
of free theories.

As a representative example we consider a theory with a polynomic
potential $\lambda V(\phi^{\dagger} \phi)$. If this interaction
term can be treated perturbatively, the Hilbert space of states
of the theory can be still expressed as a tensor product of the
Hilbert spaces of single harmonic modes, as in the free field
theory case. Therefore, the same rules we have used to assign
classical values to each field operator in the free theory can
still be applied. The only difference is  that in this case the
hamiltonian that appears in the definition of time dependent
operators contains the interaction term.

Hence, the only issue to be addressed here is whether PCP's obey
the appropriately modified classical equations of motion. In order
to settle this we consider the operator equations that the field
operator $\Phi(\vec{x},t)$ and its conjugate momentum $\Pi({\vec
x},t)$ obey
\begin{eqnarray}
\label{intoperators}
 \frac{d\Pi({\vec x},t)}{dt}&=&-\Phi({\vec x},t) -
\frac{\delta V}{\delta\Phi^*}(\Phi^* \Phi) \nonumber \\
\frac{d\Phi({\vec x},t)}{dt}&=& \Pi({\vec x},t).
\end{eqnarray}
These operator equations guarantee that classical values
$(\Pi({\vec x},t))_{cl}$ and $(\Phi({\vec x},t))_{cl}$  assigned
in the vacuum or other quantum states (as explained in the
previous subsection) obey the same classical equations of motion
on each PCFC.

\section{Pseudoclassical Paths vs. Quantum Coherent States}

Quantum coherent states are considered as the best description of
a quantum system whose dynamics approximately follows a classical
trajectory, because the quantum uncertainty is minimal on such
states as their wave function is not dispersed during its
evolution and, in addition, the average values of the position
and conjugate momentum operators evolve according to classical
equations of motion. In this sense, each coherent state can be
identified, within the precision bounds imposed by the
uncertainty principle, with a certain classical trajectory 
(see, for example, the references given in \cite{negele}).

In this section we describe the quantum coherent states of the
harmonic oscillator using the formalism of PCP's that we have
developed. This exercise will give us the opportunity to compare
the newly introduced concept of pseudoclassical paths to the
established concept of coherent quantum states. We will show that
a coherent state cannot be identified with a single PCP, and that
a single PCP cannot describes any (coherent or not) quantum state.
A whole set of PCP's with their corresponding probability
distribution is the only formal object that can describe in our 
formalism the dynamics of any given quantum state.

Let start with a brief review of the coherent quantum states of a
single harmonic oscillator. As we did in section {\bf V}, we
consider the Hilbert space generated by $a^{\dagger}_R$: $|J1
\rangle = \frac{1}{\sqrt{J1!}} (a^{\dagger}_R)^{J1} |0 \rangle$,
where $J1=0,1,...$ is a natural number. For the sake of
simplicity we can take $\kappa=1$, and then the hamiltonian of
the system is $H_R=\frac{1}{2}(Q_R^2+P_R^2)$.

Coherent states are the eigenstates of the annihilation operator
$a_R= \frac{1}{\sqrt{2}}(Q_R+iP_R)$:
\begin{equation}
\label{keypointcoherent} a|z \rangle = z |z \rangle,
\end{equation}
where $z$ is a complex number. The spectrum of the operator $a_R$
is the whole complex plane. The normalized eigenstate can be
expressed in the basis of stationary states:
\begin{equation}
|z \rangle = e^{-|z|^2/2} \sum_{J1=0}^{\infty}
\frac{z^{J1}}{\sqrt{J1!}}|J1 \rangle,
\end{equation}
and its wavefunction in the $\{|q_R \rangle \}$ representation is
given by the expression,
\begin{equation}
\label{packetcoherent} \Psi_z(q_r)=\frac{1}{\pi^{1/4}}
exp\left(-\frac{1}{2}(|z|^2-z^2)\right)
exp\left(-\frac{1}{2}(q_R-\sqrt{2}z)^2 \right), \hspace{0.7in}
q_R \in R.
\end{equation}

In particular, the vacuum state of the harmonic oscillator, $|0
\rangle$, is the coherent state that corresponds to the
eigenvalue $z=0$ of the annihilation operator: $a_R|0 \rangle=0$.

Although coherent states are not eigenstates of the hamiltonian
$H_R$, their time evolution closely resembles that of stationary
states. In the Schroedinger picture:
\begin{equation}
|z,t \rangle = exp(-i H_R t)|z \rangle = e^{-i t/2}|z e^{-i t}
\rangle,
\end{equation}
where $|z e^{-i t} \rangle$ is the coherent state that
corresponds to the eigenvalue $z'=z e^{-i t}$ of the annihilation
operator: $a_R|z e^{-i t} \rangle=z e^{-i t}|(z e^{-i t})
\rangle$. Simple algebra shows that
\begin{equation}
|\Psi_z(q_r;t)|^2 = \frac{1}{\sqrt{\pi}} exp \left([q_R - \langle
z,t|Q_R|z,t \rangle]^2 \right),
\end{equation}
the probability density of the wavepacket evolves coherently,
centered at $\langle z,t|Q_R|z,t \rangle$ without dispersion, and
the quantum uncertainty is minimal: $(\Delta Q_R) \cdot (\Delta
P_R) = \frac{1}{2}$.

As the average values of the position and momentum operators obey
classical equations of motion:
\begin{equation}
\label{CEMQ} \frac{d\langle z,t|Q_R|z,t \rangle}{dt} = \langle
z,t|P_R|z,t \rangle
\end{equation}
\begin{equation}
\label{CEMP} \frac{d\langle z,t|P_R|z,t \rangle}{dt} =-\langle
z,t|Q_R|z,t \rangle,
\end{equation}
with initial conditions
\begin{equation}
\langle z|Q_R|z \rangle = \sqrt{2} {\it Re}(z),
\end{equation}
\begin{equation}
\langle z|P_R|z \rangle = \sqrt{2} {\it Im}(z),
\end{equation}
the quantum state is considered as the closest description of a
quantum state that approximately follows a classical motion.

We would like to point out that in spite of this suggestive
interpretation the similarity between the dynamics of the
coherent quantum state and that of a classical trajectory is
limited. For example, the average value of the hamiltonian in the
quantum state $|z \rangle$,
\begin{equation}
\langle z|H_R|z \rangle = \frac{1}{2}(1 + \langle z|Q_R|z
\rangle^2 + \langle z|P_R|z \rangle^2) = \frac{1}{2}(1 + 2|z|^2),
\end{equation}
does not match the energy of the corresponding classical
trajectory,
\begin{equation}
E_{cl} = \frac{1}{2}(\langle z|Q_R|z \rangle^2 + \langle z|P_R|z
\rangle^2) = |z|^2,
\end{equation}
although the agreement does become relatively better for a large
$|z|$.

Let us now examine the coherent state $|z \rangle$ using the
formalism of PCP's. The quantum state $|z \rangle$, as any other
quantum state, corresponds to an infinite set of PCP's, each
labeled by a real number $q_R$ and whose probability to occur at
any given time is given by the squared modulus of the wavefunction
(\ref{packetcoherent}). As we know from
eq.(\ref{operators}) the time-dependent values $q_R(t)$ and $p_R(t)$
assigned to the operators $Q_R$ and $P_R$ on each of these PCP's
obey classical equations of motion. In addition, 
eq.(\ref{keypointcoherent}) implies that
\begin{equation}
\label{initialconditioncoherent} P_R|z \rangle =
i(Q_R-\sqrt{2}z)|z \rangle,
\end{equation}
which fixes the initial conditions on each PCP:
\begin{eqnarray}
\label{initialcondition} q_R(t=0) &=& q_R  \nonumber \\
 p_R(t=0) &=&i(q_R-\sqrt{2}z).
\end{eqnarray}

As an additional example of how the formalism works we compute
the value of the hamiltonian operator on each of these PCP's. First,
we express the operator $H_R$ acting on $|z \rangle$ in terms of some 
function of the operator $Q_R$ acting on $|z \rangle$:
\begin{equation}
H_R|z \rangle = \frac{1}{2}(2\sqrt{2} z Q_R - 2 z^2 +1)|z \rangle.
\end{equation}
Then, using this expression we assign to the PCP labeled by $q_R$ the 
classical value for the energy
\begin{equation}
\label{pathenergy} h(q_R)=\frac{1}{2}(2\sqrt{2} z q_R - 2 z^2 +1).
\end{equation}
As we have already noticed before, the weighted average value of
$h(q_R)$ on the whole set of PCP's is equal to $\langle z|H_R|z
\rangle$. Let us recall that classical values assigned to
operators $O$ that commute with the hamiltonian (in particular
$H_R$ itself) do not depend on time since $O(t)=e^{i H_R t} O
e^{-i H_R t} = O$.

In the context of this discussion we want to emphasize that
coherent states, as other quantum states, are described by a whole
set of PCP's and, therefore, cannot be identified with a single
PCP. Furthermore, a single PCP cannot be identified with a
particular quantum state. Let us consider, for example, a pair of
different coherent states $|z_1 \rangle$ and $|z_2 \rangle$, and
consider now the PCP labeled by a certain $q_R$ in each of these
two quantum states: we immediately see from eq.
(\ref{initialcondition}), or eq. (\ref{pathenergy}) that the two
paths do not describe the same classical trajectory.

The formalism of PCP's should be understood as a formal
construction that allows to describe the dynamics of quantum
states, coherent as well as non-coherent, in terms of
pseudoclassical stochastic events as the paths do not interfere
at all between themselves. In addition, when the formalism is
extended to describe quantum field states in terms of stochastic
classical field configurations, the usual algebraic relations
between classical observables are approximately recovered for
collective quantum operators.

Such a description is not, in general, possible in terms of
coherent states, although any quantum state of the harmonic
oscillator can be written as a linear superposition of such
coherent states, $|\Psi \rangle=\int dz \Psi(z) |z \rangle$.
After having identified each coherent state $|z \rangle$ with a
certain classical trajectory we could be tempted to describe the
generic state $|\Psi \rangle$ as a superposition of classical
trajectories as we did with the PCP's. Nevertheless, coherents
states, unlike PCP's, are actually  quantum states that do
interfere between themselves and, therefore, the average value of
a generic operator $O$ in the state $|\Psi \rangle$, $\langle
\Psi|O|\Psi \rangle$, cannot be obtained as the average of its
values on each of the coherent trajectories, $\int dz |\Psi(z)|^2
\langle z|O|z \rangle$.

Furthermore, even in the limit when coherent trajectories can be
approximately considered as non-interfering trajectories, each
with a given probability $\sim |\Psi(z)|^2$ to be realized, the
usual algebraic relations between classical observables are not
necessarily restored as we will show.

In fact, coherent states are not orthogonal to each other:
\begin{equation}
\langle z|z' \rangle = e^{-(|z|^2+|z'|^2)/2} \sum_{J1=0}^{\infty}
\frac{(z^* z')^{J1}}{J1!} = e^{-(|z|^2+|z'|^2)/2} e^{z^* z'}.
\end{equation}
Therefore, given a certain operator $O$ expressed in normal ordered form
$O(a_R^{\dagger},a_R)$ it is immediate to obtain, using eq.
(\ref{keypointcoherent}), that
\begin{equation}
\langle z|O(a_R^{\dagger},a_R)|z' \rangle = O(z^*,z') \langle
z|z' \rangle = O(z^*,z') e^{-(|z|^2+|z'|^2)/2} e^{z^* z'},
\end{equation}
and then,
\begin{equation}
\label{coherentinterference} |\langle z|O(a_R^{\dagger},a_R)|z'
\rangle| = |O(z^*,z')| e^{-(({\it Re}(z)-{\it Re}(z'))^2+({\it
Im}(z)-{\it Im}(z'))^2)/2}.
\end{equation}
This last equation means that for a general polynomial operator
$O$ only coherent states corresponding to eigenvalues $z$,$z'$
whose real and imaginary parts are significantly different can be
approximately considered as non-interfering.

When we consider the Fock space of more than one harmonic mode, as
in quantum field theory, 
the multi-mode coherent states are obtained as tensor
products of single-mode coherent states. Each coherent state is
then labeled by a sequence $\{z_{\vec n}\}$ of complex numbers,
each corresponding to the eigenvalue of the annihilation operator of a 
particular single harmonic mode in the coherent state. This set of states 
spans the multi-mode Fock space although the coherent states are not 
orthogonal to each other. From the generalization of  
eq.(\ref{coherentinterference}) to multi-mode coherent states:
\begin{equation}
\label{multicoherentinterference} |\langle \{z_{\vec
n}\}|O(a_{R,{\vec n}}^{\dagger},a_{R,{\vec n}})|\{z'_{\vec n}\}
\rangle| = |O(z_{\vec n}^*,z'_{\vec n})| e^{-\sum_{\vec n}(({\it
Re}(z_{\vec n})-{\it Re}(z'_{\vec n}))^2 + ({\it Im}(z_{\vec
n})-{\it Im}(z'_{\vec n}))^2)/2},
\end{equation}
it is clear that the exponential suppression factor is stronger
the larger the number of harmonic modes involved. In this limit,
the classical trajectories associated with each coherent state do
not interfere much between themselves.

Once we have identified this set of slightly interfering coherent states,
each labeled by a sequence of complex numbers $\{z_{\vec n}\}$, every
observable $O(t)$ can be assigned on each of them the
time-dependent value  $\langle \{z_{\vec n}\}|O(t)|\{z_{\vec n}\}
\rangle$, as we did in eqs.(\ref{CEMQ}),(\ref{CEMP}) for the
position and conjugate momentum operators. Then we could try to use this 
basis to interpret any other quantum state in terms of 
a superposition of classical trajectories.

Let us consider, in particular, the operator $\Pi({\vec x})$
introduced in eq. (\ref{momentum}). As we have noted, the
operators $P_{\vec n}$ can be expressed in terms of two hermitian
components and, therefore, for each given field theoretic coherent
state the index ${\vec n}$ indeed labels a pair $(z_{r,{\vec
n}},z_{i,{\vec n}})$ of complex numbers. The ``classical" value of
$\Pi({\vec x})$ on a coherent state trajectory is given by
\begin{equation}
\label{clmomatcoherent} \langle \{z_{\vec n}\}|\Pi({\vec
x})|\{z_{\vec n}\} \rangle = \sum_{\vec n} \frac{\kappa_{\vec
n}^{1/2}}{X^{3/2}} ({\it Im}(z_{r,{\vec n}}) + i {\it
Im}(z_{i,{\vec n}})) e^{2\pi i {\vec x} \cdot {\vec n}/X},
\end{equation}
which, in particular, vanishes for the free vacuum $\langle
0|\Pi({\vec x})|0 \rangle = 0$.
On the other hand, the value for the operator $\langle \{z_{\vec
n}\}|\Pi^{\dagger}({\vec x}) \Pi({\vec x})|\{z_{\vec n}\}
\rangle$ can be computed,
\begin{equation}
\langle \{z_{\vec n}\}|\Pi^{\dagger}({\vec x}) \Pi({\vec
x})|\{z_{\vec n}\} \rangle - \langle \{z_{\vec
n}\}|\Pi^{\dagger}({\vec x})|\{z_{\vec n}\} \rangle \langle
\{z_{\vec n}\}|\Pi({\vec x})|\{z_{\vec n}\} \rangle = \sum_{\vec
n} \frac{\kappa_{\vec n}} {X^3},
\end{equation}
for any coherent state.
We can see that for the vacuum or other coherent states with low
$z$, conditions (\ref{second}) and (\ref{condition}) are not satisfied, 
even in the case of collective observables. These conditions, which 
express the recovery in the quantum formalism of the usual algebraic 
relations between classical observables, are satisfied in the formalism
of coherent states only for states with large $z$, which correspond to 
highly populated states. 

Therefore, the formalism of coherent states does not permit an adequate 
description of the dynamics of quantum fluctuations of physical 
observables in scarcely populated states in terms of classical 
trajectories. For example, the free vacuum state is itself the coherent 
state corresponding to the set of eigenvalues ${z_{\vec k}=0}$, 
or in other words, $|0 \rangle = \int dz \delta(z) |z \rangle$. Such a 
state would correspond in the description of coherent states to a single 
classical harmonic trajectory with initial conditions $\langle Q 
\rangle = \langle P \rangle = 0$, which corresponds to a 
static trajectory. 
In contrast, in the formalism of PCP's the free vacuum is 
represented as a linear superposition of an infinite number of randomly 
distributed paths, which we have used to describe the dynamics of the
quantum fluctuations of physical observables, and conditions 
(\ref{second}) and (\ref{condition}) 
are recovered on each "pseudoclassical" trajectory for collective 
observables, as we have proved. 

Let us recall that in the PCP's formalism each classical path
is labeled by a set of eigenvalues of a complete representation of
commuting observables, for example, the $\{Q\}$ representation, but none 
of these paths can be understood as a quantum state. In
the formalism of coherent states, on the other hand, each classical 
trajectory corresponds to a coherent quantum state which is labeled by the 
eigenvalues of a set of annihilation operators, which are not hermitian 
operators and, therefore, do not correspond to physical observables.

Finally, an additional comment on the concept of classicality
that we have introduced together with the concept of PCP's (or
PCFC's for quantum field theoretic states) is pertinent. We have
found that the pseudoclassical values on PCP's assigned to
certain class of observables, which we have identified as
collective observables, obey approximately the standard algebraic
relations between classical observables. The dynamics of the
quantum system can then be approximately described, as far as
these observables are concerned, as a classical statistical
system. It should be stressed, though, that the indeterminacy of
the system has a quantum character. In the formalism of PCP's  we
identify  classical dynamics associated with a certain class of
observables, whereas in the formalism of coherent states
classicality is associated with a certain class of quantum states.

\section{Summary}

We have developed a formalism in which the quantum states of weakly
interacting systems can be described as a linear combination of incoherent paths
which obey classical equations of motion with
constrained initial conditions and have definite probabilities
to be realized due to quantum fluctuations. In this formalism we may
represent the dynamics of quantum fluctuations of collective operators
which depend on many degrees of freedom in terms of
unconstrained classical stochastic processes
and the concept of classical dynamics of quantum fluctuations can be
defined. Classicality and decoherence of quantum fluctuations are viewed
as collective effects.

Although we have focused on a complex relativistic scalar field
the whole formalism, and in particular the  expression for the
random fluctuations of the energy contained in a finite volume,
can be generalized with minor changes to describe quantum
fluctuations of gauge bosons, fermions or non-relativistic fields
with the condition that the independent degrees of freedom are
only weakly coupled.

The formalism is based on the following  basic
properties of weakly coupled
quantum field theories:

1) the Hilbert space of weakly interacting theories is the tensor
product of the spaces of single modes states

2) Lorentz covariant collective observables can be expressed as a
direct sum of independent single-mode operators that are
necessarily causally disconnected at some initial time $t=0$
whatever the actual quantum state of the theory is;

3) the canonical operators $Q,P$ in the Heisenberg picture obey
classical equations of motion.

\section{Acknowledgments}

We warmly acknowledge very useful discussions with J. Oaknin. We
wish to thank D. Cohen for discussions and comments, and S. de
Alwis for drawing our attention to \cite{GMH}. This research
is supported by grant no. 174/00-2 of the Israel Science Foundation.

\end{document}